\documentclass[a4paper]{amsart}
\usepackage{latexsym}
\usepackage{amsmath}
\usepackage{amssymb}
\usepackage{amsfonts}
\usepackage{stmaryrd}
\usepackage{bbm}

\usepackage{color}
\definecolor{Blue}{rgb}{0.,0.,1.}
\definecolor{Red}{rgb}{1.,0.,0.}

\usepackage[T1]{fontenc}

\def\init{\setcounter{equation}{0}}

\newtheorem{theoreme}{Theorem }[section]
\newtheorem{proposition}[theoreme]{Proposition}
\newtheorem{lemma}[theoreme]{Lemma}
\newtheorem{definition}[theoreme]{Definition}

\newtheorem{remark}[theoreme]{Remark}

\newcommand\bbone{\ensuremath{\mathbbm{1}}}

\def\rr{\mathbb{R}}
\def\cc{\mathbb{C}}
\def\nn{\mathbb{N}}

\def\one{\bbone}

\newcounter{smallarabics}
\newenvironment{arabicenumerate}
{\begin{list}{{\normalfont\textrm{(\arabic{smallarabics})}}}
  {\usecounter{smallarabics}\setlength{\itemindent}{0cm}
   \setlength{\leftmargin}{5ex}\setlength{\labelwidth}{4ex}
   \setlength{\topsep}{0.75\parsep}\setlength{\partopsep}{0ex}
   \setlength{\itemsep}{0ex}}}
{\end{list}}

\newcounter{smallroman}

\newcommand{\ben}{\begin{arabicenumerate}}  
\newcommand{\een}{\end{arabicenumerate}}

\def\e{{\rm e}}

\def\i{{\rm i}}
\def\d{{\rm d}}
\def\12{\frac{1}{2}}
\def\cinf{C^{\infty}}

\def\proof{\noindent{\bf  Proof. }}

\def\coinf{C_{0}^{\infty}}

\def\qed{$\Box$}

\def\cH{{\mathcal H}}

\def\cK{{\mathcal K}}

\def\K{{\mathcal K}}

\def\ch{{\mathfrak h}}

\def\p{\partial}
\def\s{{\rm s}}

\def\x{\langle x\rangle}

\def\zbar{\bar{z}}

\def\pfi2{P(\varphi)_{2}}

\newcommand{\beq}{\begin{equation}}
\newcommand{\eeq}{\end{equation}}
\newcommand{\bet}{\begin{theoreme}}
\newcommand{\eet}{\end{theoreme}}
\newcommand{\bel}{\begin{lemma}}
\newcommand{\eel}{\end{lemma}}
\newcommand{\bep}{\begin{proposition}}
\newcommand{\eep}{\end{proposition}}
\newcommand{\bear}[1]{\begin{array}{#1}}
\newcommand{\ear}{\end{array}}

\begin{document}

\def\triple{\interleave}
\def\Gh{\Gamma(\ch)}
\def\Dom{{\rm Dom}}
\def\y{\langle y\rangle}
\def\rx{{\rm x}}
\def\ry{{\rm y}}

\title{Infrared  problem  \\ for  the Nelson model on static space-times}
\author{C. G\'erard}
\address{D\'epartement de Math\'ematiques, Universit\'e de Paris XI, 91405 Orsay Cedex France}
\email{christian.gerard@math.u-psud.fr}
\author{F. Hiroshima}
\address{Department of Mathematics, University of Kyushu, 6-10-1, Hakozaki, Fukuoka, 812-8581, Japan}
\email{hiroshima@math.kyushu-u.ac.jp}
\author{A. Panati}
\address{PHYMAT, Universit\'e Toulon-Var 83957 La Garde Cedex France}
\email{annalisa.panati@univ-tln.fr}
\author{A. Suzuki}
\address{Department of Mathematics, Faculty of Engineering, Shinshu University, 4-17-1 Wakasato, Nagano 380-8553, Japan}
\email{sakito@math.kyushu-u.ac.jp}
\begin{abstract}
We consider  the Nelson model on some static space-times
and investigate the problem of existence of a ground state. Nelson models with variable
coefficients arise when one replaces in the usual Nelson model 
the flat Minkowski metric by a  static metric, allowing also the
boson mass to depend on position.   We investigate the
existence of a ground state of the Hamiltonian in the presence of the
infrared problem, i.e. assuming that the boson  mass  $m(x)$ tends to $0$ at spatial 
infinity. We show that if $m(x)\geq C |x|^{-1}$ at infinity for some $C>0$ then the Nelson Hamiltonian has a ground state.
\end{abstract}
\date{November 2010}
\maketitle

\section{Introduction}\label{intro}\init

The study of  Quantum Field Theory on curved space-times  has seen  important developments since the seventies. Probably the most spectacular prediction in this domain is the {\em Hawking effect} \cite{Ha,  FH, Ba},  predicting that a star collapsing to a black hole asymptotically emits a thermal radiation.
A related effect is the {\em Unruh effect} \cite{Un, Un-W, dB-M},  where an accelerating observer in Minkowski space-time sees the vacuum state as a thermal state. 

Another important development is the  use of {\em microlocal analysis} to study  free or quasi-free states on globally hyperbolic space-times, 
which started with the seminal work by Radzikowski \cite{Ra1, Ra2} , who proved  that Hadamard states (the natural substitutes for vacuum states on  curved space-times) can be characterized in terms of microlocal properties of their two-point functions.  The use of microlocal analysis in this domain was further developed for example in \cite{BFK}, \cite{Sa}.

Most of these works  deal with  free or quasi-free states, because of the well-known difficulty to construct an interacting, relativistic quantum field theory, even on Minkowski space-time.  

However in recent years a lot of effort was devoted to the  rigorous 
study of {\em interacting} non-relativistic models on Minkowski space-time, typically obtained by coupling a relativistic quantum field to non-relativistic particles. 
The two main examples are  {\em non-relativistic QED}, where the quantized Maxwell field is minimally coupled to a non-relativistic particle  and the {\em Nelson model}, where a scalar bosonic field is linearly coupled to a non-relativistic particle. For both models it is necessary to add an ultraviolet cutoff in the interaction term to rigorously construct the associated Hamiltonian.

In both cases the models can be constructed on a Fock space with relatively little efforts, and several properties of the quantum Hamiltonian  $H$ can be 
rigorously studied.  One of them, which will also be our 
main interest  in this paper, is the question of the {\em existence of 
a ground state}.  Obviously the fact that $H$ has a ground state is an
important physical property of the Nelson model. For example a consequence of the existence of a ground state is that 
{\em scattering states} can quite easily be constructed. These states
describe   the ground state of $H$ with a finite number of  additional asymptotically free bosons.

When  $H$ has no ground state one usually speaks of the {\em infrared problem} or {\em infrared divergence}.  The infrared problem arises when the
emission probability of bosons becomes infinite with increasing wave length.  If the infrared problem occurs,  the scattering theory has to be modified: all scattering states contain an infinite number of low energy (soft) bosons (see eg \cite{DG1}).
Among many  papers  devoted to this question, let us mention 
\cite{AHH, BFS, BHLMS, G, H, LMS, Sp} for  the   Nelson model, and \cite{GLL} for non-relativistic QED.

Our goal in this paper is to study the existence of a ground 
state for the Nelson model on a {\em static} space-time, allowing also for a position-dependent mass.  This model is  obtained by linearly coupling the Lagrangians of a Klein-Gordon field   and of a non-relativistic particle on a static space-time (see Subsect.  \ref{sec00.2}).
 We believe that this model, although non-relativistic, 
 is an interesting testing ground for  the generalization of results for free or quasi-free models on curved space-times 
 to some interacting  situations. Let us also mention that for the Nelson model on Minkowski space-time 
 the removal of the ultraviolet cutoff  can be done by relatively easy arguments. After removal of the ultraviolet cutoff, the Nelson model becomes a
 {\em local} (although non-relativistic) QFT model.   In a subsequent paper \cite{GHPS3}, we will  show that the ultraviolet cutoff can be removed for the Nelson model on a static space-time.
 
 Most of our discussion will be focused on the role of the variable mass term on the ground state existence. Note that when one considers a massive  Klein-Gordon field in the Schwarzschild metric, the effective mass tends to $0$ at the black hole horizon (see eg \cite{Ba}).    We believe that the study of the Nelson model with a variable mass vanishing at spatial infinity will be a first step  towards the extension of the rigorous justification of the Hawking effect in \cite{Ba} to some interacting models.

\subsection{The  Nelson model on Minkowski space-time}\label{intro.1}
In this subsection we quickly describe the usual Nelson model on Minkowski space-time.
The {\em Nelson} model  describes a  scalar  bosonic field linearly coupled to a  quantum mechanical particle. It is formally defined by the  Hamiltonian
\[
H=\12 p^{2}+ W(q)+ \12\int_{\rr^{3}} \pi^{2}(\rx)+(\nabla\varphi(\rx))^{2} + m^{2}\varphi^{2}(\rx)\d \rx + \int_{\rr^{3}} \varphi(\rx)\rho(\rx-q)\d \rx,
\]
where $\rho$ denotes a cutoff function, $p$, $q$ denote the position and momentum of the particle, $W(q)$ is an external potential
 and  $\varphi(\rx)$, $\pi(\rx)$ are the canonical 
field position and momentum. 

The Nelson model arises from the quantization of the following coupled Klein-Gordon and Newton system:
\begin{equation}
\label{e1.intro}
\left\{
\begin{array}{l}
(\Box+ m^{2})\varphi(t, \rx)= -\rho(\rx- q_{t}), \\[2mm]
\ddot{q}_{t}= -\nabla_{q}W(q_{t})- \int \varphi(t, \rx)\nabla_{\rx}\rho(\rx- q_{t})\d \rx,
\end{array}
\right.
\end{equation}
were $\Box$ denotes the d'Alembertian on  the Minkowski space-time $\rr^{1+3}$. 
The cutoff function $\rho$ plays the role of  an ultraviolet cutoff and amounts to replacing the  quantum mechanical  point particle by a charge density.

To distinguish the Nelson model on Minkowski space-time from its generalizations  that will be described later in the introduction, we will call it the 
{\em usual} (or {\em  constant coefficients}) {\em Nelson model}.

For the usual Nelson model the situation is as follows:
one assumes a stability condition (see Subsect. \ref{non-confining}), 
implying that states with energy close to the bottom of the spectrum are localized in the particle position. 
Then if the bosons are massive i.e. if $m>0$ $H$ has a ground state (see
eg \cite{G}). On the contrary if $m=0$  and  $\int \rho(x)\d x\neq 0$
then $H$ has no ground state (see \cite{DG1}).

\subsection{The Nelson model with variable coefficients}\label{intro.2}
We now describe   a generalization of the usual Nelson model, obtained by replacing  the free Laplacian $-\Delta_{\rx}$ by a  general second order differential operator  and the constant mass term $m$ by a function $m(\rx)$.  We set:
\[
h:= -\sum_{1\leq j,k\leq d} c(\rx)^{-1}\p_{j}a^{jk}(\rx)\p_{k}c(\rx)^{-1}+ m^{2}(\rx), 
\]
for a Riemannian metric $a_{jk}\d\rx^{j}\d\rx^{k}$ and two functions $c(\rx)$, $m(\rx)>0$, 
and consider the generalization of (\ref{e1.intro}):
\beq\label{e2.intro}
\left\{
\begin{array}{l}
\p_{t}^{2}\phi(t,\rx) +h \phi(t, \rx)+ \rho(\rx- q_{t})=0,\\[2mm]
\ddot{q}_{t}= -\nabla_{\rx}W(q_{t})-\int_{\rr^{3}} \phi(t,\rx)\nabla_{\rx}\rho(\rx -q_{t})|g|^{\12}\d^{3}\rx.
\end{array}
\right.
\eeq
Quantizing the field equations (\ref{e2.intro}), we obtain a  Hamiltonian $H$ acting on  the Hilbert space 
$L^{2}(\rr^{3})\otimes \Gamma_{\rm s}(L^{2}(\rr^{3}))$ (see  Sect. \ref{sec2}), which we call a {\em Nelson Hamiltonian with variable coefficients}. Formally $H$ is defined by the following expression:
\begin{align}
H=&\12 p^{2}+ W(q)\label{e3.intro}\\[2mm]+ &\12\int_{\rr^{3}} \pi^{2}(\rx)
+\sum_{jk}\p_{j}\big(c(\rx)^{-1}\varphi(\rx)\big)a^{jk}(\rx)\big(\p_{k}c(\rx)^{-1}\varphi(\rx)\big)+ m^{2}(\rx)\varphi^{2}(\rx)\d \rx\notag\\[2mm]
+& \int_{\rr^{3}} \varphi(\rx)\rho(\rx-q)\d \rx.\notag
\end{align}
The main example of a variable coefficients Nelson model is obtained by replacing  in the usual Nelson model
the flat Minkowski metric on $\rr^{1+3}$ by a {\em static} Lorentzian metric, and by allowing also the mass $m$ to be position dependent. 
Recall that  a static metric on $\rr^{1+3}$ is of the form
\[
g_{\mu\nu}(x)\d x^{\mu}\d x^{\nu}= -\lambda(\rx)\d t \d t+
\lambda(x)^{-1}h_{\alpha\beta}(\rx)\d\rx^{\alpha}\d\rx^{\beta},
\]
where $x=(t, \rx)\in \rr^{1+3}$, $\lambda(\rx)>0$ is a smooth function, and $h_{\alpha, \beta}(\rx)$ is a
 Riemannian metric on $\rr^{3}$. We show in Subsect. \ref{sec00.3} that  the natural Lagrangian for a point 
particle coupled to a scalar field  on $(\rr^{1+3}, g)$ leads (after a change of field variables) 
to  the  system  (\ref{e2.intro}). 
\subsection{The infrared problem}\label{intro.3}
Assuming reasonable hypotheses on the matrix \\$[a^{jk}](\rx)$ and the functions $c(\rx)$, $m(\rx)$ it is easy to see that the formal expression (\ref{e3.intro}) can be rigorously defined as a bounded below selfadjoint operator $H$.  

The question we address in this paper is the problem of existence of a ground state for $H$.   Variable coefficients Nelson models are examples of an abstract class of QFT Hamiltonians called {\em abstract Pauli-Fierz Hamiltonians} (see eg \cite{G}, \cite{BD} and Subsect. \ref{sec4.1}).   If $\omega$ is  the {\em one-particle energy}, the constant $m:=\inf\sigma(\omega)$ can be called the (rest) mass of the bosonic field, and abstract Pauli-Fierz Hamiltonians fall naturally into two classes: massive models if $m>0$ and massless if  $m=0$. 

For massive models, $H$ typically has a ground state, if we assume either that the quantum particle is confined or a stability condition (see Subsect. 
\ref{non-confining}). 
In this paper we concentrate on  the massless case and hence our typical assumption will be that 
\[
\lim_{\rx\to\infty}m(\rx)=0.
\]
It follows that bosons of arbitrarily small energy may be present.    The main result of this paper is that  the existence or non-existence of a ground state for $H$ depends on the rate of decay of the function $m(\rx)$. In fact we show in Thm. \ref{mainmain} that if 
\[
m(\rx)\geq a\langle\rx\rangle^{-1}, \hbox{ for some }a>0,
\]
and if the quantum particle is confined, then $H$ has a ground state.
In a subsequent paper \cite{GHPS2}, we will show that if
\[
0\leq m(\rx)\leq C\langle \rx\rangle^{-1-\epsilon}, \hbox{ for some }\epsilon>0,
\]
then $H$ has no ground state.   Therefore Thm. \ref{mainmain} is sharp with respect to the decay rate of the mass at infinity.

(If $h= -\Delta + \lambda m^{2}(\rx)$ for $m(\rx)\in O(\langle \rx\rangle^{-3/2})$ and the coupling  constant $\lambda$ is sufficiently small the same result is shown in \cite{GHPS1}).

\subsection{Notation}
We collect here some notation for the reader's convenience.

If $x\in \rr^{d}$, we set $\langle x\rangle= (1+ x^{2})^{\12}$.

The domain of a linear operator $A$ on some Hilbert space $\cH$ will be denoted by $\Dom A$, and its spectrum by $\sigma(A)$.

If $\ch$ is a Hilbert space, the {\em bosonic Fock space} over $\ch$ denoted by $\Gamma_{\rm s}(\ch)$ is
\[
\Gamma_{\rm s}(\ch):=\bigoplus_{n=0}^{\infty}\otimes_{\rm s}^{n}\ch.
\]
We denote by  $a^{*}(h)$, $a(h)$ for $h\in \ch$ the {\em creation/annihilation operators} acting on $\Gamma_{\rm s}(\ch)$. The (Segal) {\em field operators} 
$\phi(h)$ are defined as $\phi(h):=\frac{1}{\sqrt{2}}(a^{*}(h)+ a(h))$. 

If $\cK$ is another Hilbert space and $v\in B(\cK, \cK\otimes\ch)$, then one defines  the operators $a^{*}(v), \ a(v)$ as unbounded operators on $\cK\otimes \Gamma_{\rm s}(\ch)$ by:
\[\begin{array}{l}
a^*(v)\Big|_{\cK\otimes\bigotimes_\s^n\ch}
:=\sqrt{n+1}\Big(\one_\cK\otimes {\mathcal S}_{n+1}\Big)
\Big(v\otimes\one_{\bigotimes_\s^n\ch}\Big),\\[3mm]
a(v):=\big(a^*(v)\big)^*,\\[3mm]
\phi(v):=\frac{1}{\sqrt2}(a(v)+a^*(v).
\end{array}\]
 They satisfy the estimates
\beq
\|a^{\sharp}(v)(N+1)^{-\12}\|\leq \|v\|,
\label{ju88}
\eeq
where $\|v\|$ is the norm of $v$ in $B(\K,\K\otimes\ch)$.

If $b$ is a selfadjoint operator on $\ch$ its second quantization $\d \Gamma(b)$ is defined as:
 \[
d\Gamma(b)\Big|_{\bigotimes_\s^n\ch}:
=\sum\limits_{j=1}^n\underbrace{\one\otimes\cdots\otimes\one}_{j-1}
\otimes b\otimes \underbrace{\one\otimes\cdots\otimes\one}_{n-j}.
\]

\section{The Nelson model on static space-times}\init\label{sec00}
In this section we discuss the Nelson model on static space-times, which is the main example 
of  Hamiltonians that will be studied in the rest of the paper. It is
convenient to start with the Lagrangian framework.
\subsection{Klein-Gordon equation on static space-times}\label{sec00.1}
Let $g_{\mu\nu}(x)$ be a Lor\-entz\-ian metric of signature $(-, +,+,+)$
on $\rr^{1+3}$. Set $|g|={\rm det}[g_{\mu\nu}]$, $[g^{\mu
\nu}]= [g_{\mu\nu}]^{-1}$. 
Consider the Lagrangian
\[
L_{\rm free}(\phi)(x)= \12\p_{\mu}\phi(x) g^{\mu\nu}(x)\p_{\nu}\phi(x)+ \12 m^{2}(x) \phi^{2}(x),
\]
 for a function $m:\rr^{4}\to \rr^{+}$ and the associated action:
\[
S_{\rm field}(\phi)= \int_{\rr^{4}}L_{\rm free}(\phi)(x)|g|^{\12}(x)\d^{4}x,
\]
where $\phi: \rr^{4}\to \rr$. 
The Euler-Lagrange equations yield the {\em Klein-Gordon equation}:
\[
\Box_{g}\phi + m^{2}(x)\phi=0,
\]
for
\[
\Box_{g}= -|g|^{-\12}\p_{\mu}|g|^{\12}g^{\mu\nu}\p_{\nu}.
\]
Usually one has
\[
\12 m^{2}(x)=\12 (m^{2}+ \theta R(x)),
\]
where  $m\geq 0$ is the mass and $R(x)$ is the
scalar curvature of the metric $g_{\mu\nu}$, (assuming of course that the function on the right is positive). In particular 
   if $m=0$ and  $\theta=\frac{1}{6}$ one obtains the so-called conformal wave equation. 
   
We set $x= (t, \rx)\in \rr^{1+3}$.  The metric $g_{\mu\nu}$ is {\em static} if:
\[
g_{\mu\nu}(x)\d x^{\mu}\d x^{\nu}= -\lambda(\rx)\d t \d t+
\lambda(x)^{-1}h_{\alpha\beta}(\rx)\d\rx^{\alpha}\d\rx^{\beta},
\]
where $\lambda(\rx)>0$ is a smooth function and  $h_{\alpha\beta}$ is a Riemannian metric on $\rr^{3}$.   We assume also that 
$m^{2}(x)= m^{2}(\rx)$ is independent on $t$.

Setting $\phi(t, \rx)=\lambda |h|^{-1/4}
\tilde{\phi}(t, \rx)$, we obtain that $\tilde{\phi}(t, \rx)$ satisfies the equation:
\[
\p_{t}^{2}\tilde{\phi} - \lambda|h|^{-1/4}
\p_{\alpha}|h|^{\12}h^{\alpha\beta}\p_{\beta}|h|^{-1/4}\lambda\tilde{\phi} + m^{2}\lambda\tilde{\phi}=0.
\]
We note that $|h|^{-1/4}
\p_{\alpha}|h|^{\12}h^{\alpha\beta}\p_{\beta}|h|^{-1/4}$ is (formally) self-adjoint on $L^{2}(\rr^{3}, \d \rx)$ and is the Laplace-Beltrami operator $\Delta_{h}$associated to the 
Riemannian metric $h_{\alpha\beta}$ 
(after the usual density change $u\mapsto |h|^{1/4}u$ to work on  the
Hilbert space $L^{2}(\rr^{3}, \d \rx))$.
\subsection{Klein-Gordon field coupled to a non-relativistic
particle}\label{sec00.2}
We now couple the Klein-Gordon field to a non-relativistic particle.
We  fix a mass $M>0$, a charge density $\rho: \rr^{3}\to
\rr^{+}$ with $q=\int_{\rr^{3}}\rho(\ry)\d^{3}\ry\neq 0$ and a real potential $W: \rr^{3}\to \rr$. 
The action for the coupled system is
\[
S= S_{\rm part}+ S_{\rm field}+ S_{\rm int},
\]
for
\[
\begin{array}{l}
S_{\rm part}=\int_{\rr}\frac{M}{2}|\dot{\rx}(t)|^{2}- W(\rx(t))\d t,
\\[2mm]
S_{\rm int}= \int_{\rr^{4}} \phi(t, \rx) \rho(\rx
-\rx(t))|g|^{\12}(x)\d^{4}x.
\end{array}
\]
The Euler-Lagrange equations are:
\[
\left\{
\begin{array}{l}
\Box_{g}\phi(t,\rx) + m^{2}(t, \rx)\phi(t, \rx)+ \rho(\rx- \rx(t))=0,\\[2mm]
M\ddot{\rx}(t)= -\nabla_{\rx}W(\rx(t))-\int_{\rr^{3}} \phi(t,
\rx)\nabla_{\rx}\rho(\rx -\rx(t))|g|^{\12}\d^{3}\rx.
\end{array}
\right.
\]
Doing the same change of  field variables as in Subsect. \ref{sec00.1} and deleting the tildes, we
obtain the system:
\beq\label{nilson}
\left\{
\begin{array}{l}
\p_{t}^{2}\phi -\lambda \Delta_{h}\lambda \phi + m^{2}\lambda \phi +
\rho(\rx-\rx(t))=0,\\[2mm]
M\ddot{\rx}(t)= -\nabla W(\rx(t))- \int_{\rr^{3}}\phi(t,
\rx)\nabla\rho(\rx-\rx(t))d^{3}\rx.
\end{array}
\right.
\eeq
\subsection{The Nelson model on a static space-time}\label{sec00.3}
If the metric is static, the equations (\ref{nilson})  are clearly  Hamiltonian equations for the classical Hamiltonian 
$H= H_{\rm part}+ H_{\rm field}+ H_{\rm int}$, where:
\[
H_{\rm part}(\rx, \xi)=  \frac{1}{2M}\xi^{2}+ W(\rx),
\]
\[
\begin{array}{rl}
&H_{\rm field}(\varphi, \pi)\\[4mm]
=&\12\int_{\rr^{3}} \pi^{2}(\rx)-
\varphi(\rx)\lambda(\rx)\Delta_{h}\lambda(\rx)
\varphi(\rx)+ m^{2}(\rx)\lambda(\rx)\varphi^{2}(\rx)\d \rx,
\end{array}
\]
\[
H_{\rm int}(\rx, \xi, \varphi, \pi)= \int_{\rr^{3}} \rho(\ry-
\rx)\varphi(\ry)\d \ry.
\]
The classical phase space is as usual $\rr^{3}\times \rr^{3}\times L^{2}_{\rr}(\rr^{3})\times L^{2}_{\rr}(\rr^{3})$, with the symplectic form
\[
(\rx, \xi, \varphi, \pi)\omega(\rx', \xi', \varphi', \pi')= \rx\cdot \xi'- \rx'\cdot \xi + \int_{\rr^{3}}\varphi(\rx)\pi'(\rx)- \pi(\rx)\varphi'(\rx)\d \rx.
\]
The usual quantization scheme leads to the Hilbert space:
\[
L^{2}(\rr^{3}, \d\ry)\otimes\Gamma_{\rm s}(L^{2}(\rr^{3}, \d \rx)),
\]
where $\Gamma_{\rm s}(\ch)$ is the bosonic Fock space over the one-particle space $\ch$, and to the quantum Hamiltonian:
\[
H= (-\12\Delta_{\ry}+ W(\ry))\otimes \one + \one\otimes \d\Gamma(\omega)+\frac{1}{\sqrt{2}}\left(a^{*}(\omega^{-\12}\rho(\cdot -\ry)+ 
a(\omega^{-\12}\rho(\cdot -\ry)\right),
\]
where 
\[
\omega= (-\lambda\Delta_{h}\lambda+ m^{2}\lambda)^{\12},
\]
$\d\Gamma(\omega)$ is the usual second quantization of $\omega$ and $a^{*}(f),\ a(f)$ are the creation/an\-nihilation operators on $\Gamma_{\rm s}(L^{2}(\rr^{3}, \d \rx))$.

\section{The Nelson Hamiltonian with variable coefficients}\init\label{sec2}
In this section we define the Nelson model with variable coefficients that will be studied in the rest of the paper.  
We will deviate slightly from the notation in Sect. \ref{sec00} by denoting by $x\in \rr^{3}$ (resp. $X\in \rr^{3}$) the boson (resp. electron) position.
As usual we set $D_{x}= \i^{-1}\nabla_{x}$, $D_{X}= \i^{-1}\nabla_{X}$.
\subsection{Electron Hamiltonian}\label{sec2.1}
We define the electron Hamiltonian as:
\[
K:= K_{0}+ W(X), 
\]
where
\[
 K_{0}=\sum_{1\leq j, k\leq 3}D_{X_{j}} A^{jk}(X)D_{X_{k}},
\]
 acting on $\cK:= L^{2}(\rr^{3},\d X)$, where:
\[
(E1)\  \ C_{0}\one \leq [A^{jk}(X)]\leq C_{1}\one,\ C_{0}>0.
\]
We assume that $W(X)$ is a real potential such that $K_{0}+W$ is
essentially selfadjoint and bounded below. We denote by $K$ the
closure of $K_{0}+W$.  Later we will assume the following {\em
confinement condition }:
\[
(E2)\ \ W(X)\geq C_{0}\langle X\rangle^{2\delta}- C_{1}, \hbox{ for some }\delta>0. 
\]
Physically this condition means that the electron is confined.  
As is well known (see eg \cite{GLL}) for the question of existence  of a ground state 
, this condition can be replaced by a {\em stability condition}, meaning that states near the bottom of the spectrum of the Hamiltonian are confined in
the electronic variables by energy conservation.

We will discuss the extension of our results when one assume the stability condition in Subsect. \ref{non-confining}.

\subsection{Field Hamiltonian}\label{sec2.2}
Let:
\[
\begin{array}{rl}
h_{0}:=-&\sum_{1\leq j,k\leq d} c(x)^{-1}\p_{j}a^{jk}(x)\p_{k}c(x)^{-1},\\[2mm]
 h:=& h_{0}+ m^{2}(x),
\end{array}
\]
with $a^{jk}$, $c$, $m$  are real functions and:
\[
(B1) \ \begin{array}{l}
C_{0}\one\leq [a^{jk}(x)]\leq C_{1}\one,  \ C_{0}\leq c(x)\leq C_{1}, \ C_{0}>0, \\[2mm]
\p_{x}^{\alpha}a^{jk}(x)\in O(\langle x\rangle^{-1}), \ |\alpha|\leq 1, \ \p_{x}^{\alpha}c(x)\in O(1), \ |\alpha|\leq 2,\\[2mm]
\p_{x}^{\alpha}m(x)\in O(1), \ |\alpha|\leq 1.
\end{array}
\]
Clearly $h$ is selfadjoint on  $H^{2}(\rr^{3})$ and
$h\geq 0$. The {\em one-particle space} and 
{\em one-particle energy} are:
\[
\ch:=L^{2}(\rr^{3}, \d x), \ \omega:=h^{\12}.
\]
The constant:
\[
\inf\sigma(\omega)=:m\geq 0,
\]
can be viewed as the {\em mass} of the scalar bosons.  

The following lemma is easy;
\begin{lemma}\label{akito}
\ben
\item One has ${\rm Ker}\omega=\{0\}$,
\item
Assume in addition to  (B1) that $\lim_{x\to \infty}m(x)=0$. Then $\inf \sigma(\omega)=0$.
\een
\end{lemma}
\proof It follows from (B1) that
\[
(u |hu)\leq C_{1}(c^{-1}u | - \Delta c^{-1}u)+ (c^{-1}u|c^{-1} m^{2} u), \ u\in H^{2}(\rr^{3}).
\]
Therefore if $hu=0$ $u$ is constant. It follows also from (B1) that $c(x)^{-1}$ preserves $H^{2}(\rr^{3})$. Therefore by the variational principle
\[
m^{2}= \inf \sigma(h)\leq C_{1}\inf \sigma(-\Delta + c^{-2}(x)m^{2}(x))=0.
\]
This proves (2). \qed

The Nelson
Hamiltonian defined  below will be called {\em massive} (resp. {\em
massless}) if $m>0$ (resp. $m=0$.)
The field Hamiltonian is
\[
\d\Gamma(\omega),
\]
acting on the bosonic Fock space $\Gamma_{\rm s}(\ch)$.

\subsection{Nelson Hamiltonian}\label{sec2.3}
Let $\rho\in S(\rr^{3})$, with $\rho\geq 0$, $q=\int_{\rr^{3}} \rho(y)\d y\neq 0$.  We set:
\[
 \rho_{X}(x)= \rho(x-X)
\]
  and define the {\em UV cutoff fields} as:
\beq\label{e0.1}
\varphi_{\rho}(X):= \phi( \omega^{-\12} \rho_{X}),
\eeq
where for $f\in \ch$, $\phi(f)$ is the Segal field operator:
\[
\phi(f):= \frac{1}{\sqrt{2}}\left(a^{*}(f)+ a(f)\right).
\]
Note that setting
\[
\varphi(X):= \phi(\omega^{-\12}\delta_{X}),
\]
one has $\varphi_{\rho}(X)= \int \varphi(X-Y)\rho(Y)\d Y$.
\begin{remark}
One can think of another definition of UV cutoff fields, namely:
\[
\tilde{\varphi}_{\chi}(X):= \phi(\omega^{-\12}
\chi(\omega)\delta_{X}),
\]
for $\chi\in S(\rr)$, $\chi(0)=1$. In the constant coefficients case where $h=-\Delta$
both definitions are equivalent. In the variable coefficients case
 the natural definition (\ref{e0.1}) is much more convenient.
\end{remark}

The {\em Nelson Hamiltonian} is:
\beq\label{e2.1}
H:= K\otimes\one + \one \otimes \d\Gamma(\omega)+ \varphi_{\rho}(X),
\eeq
acting on
\[
\cH= \cK\otimes\Gamma_{\rm s}(\ch).
\]
Set also:
\[
H_{0}:=K\otimes\one + \one \otimes \d\Gamma(\omega),
\]
which is selfadjoint on its natural domain.
The following lemma is standard.
\begin{lemma}\label{0.1} Assume  hypotheses (E1), (B1). Then 
$H$ is selfadjoint and bounded below on $D(H_{0})$.
\end{lemma}
\proof it suffices to apply results on abstract Pauli-Fierz
Hamiltonians (see eg \cite[Sect.4]{GGM}). $H$ is an abstract
Pauli-Fierz Hamiltonian with coupling operator $v\in B(\cK,
\cK\otimes\ch)$ equal to:
\[
L^{2}(\rr^{3}, \d X)\ni u\mapsto \omega^{-\12}\rho(x-X)u(X)\in L^{2}(\rr^{3}, \d X)\otimes L^{2}(\rr^{3}, \d x)
\]
Applying \cite[Corr. 4.4]{GGM}, it suffices to check that
$\omega^{-\12}v\in B(\cK, \cK\otimes\ch)$. Now
\[
\|\omega^{-\12}v\|_{B(\cK, \cK\otimes\ch)}=(\sup_{X\in
\rr^{3}}\|\omega^{-1}\rho_{X}\|^{2})^{\12}
\]
Using that $h\geq C D_{x}^{2}$ and  the Kato-Heinz inequality, we obtain that
$\omega^{-2}\leq C |D_{x}|^{-2}$, hence it suffices to check that the map
\[
L^{2}(\rr^{3}, \d X)\ni u\mapsto |D_{x}|^{-1}\rho(x-X)u(X)\in L^{2}(\rr^{3}, \d X)\otimes L^{2}(\rr^{3}, \d x)
\]
is bounded, which is well known.  \qed
\section{Existence of a ground state}\label{sec4}\init
In this section we will prove  our main result about the existence of a ground state for variable coefficients Nelson Hamiltonians.  
This result will be deduced from an abstract
existence result extending the one in \cite{BD}, whose proof is outlined in Subsects. \ref{sec4.1}, \ref{sec4.2} and \ref{sec4.3}.
\begin{theoreme}\label{mainmain}
 Assume hypotheses (E1), (B1). Assume in addition that:
 \[
m(x)\geq a \x^{-1}, \hbox{ for some }a>0,
\]
and (E2) for some $\delta>\frac{3}{2}$. Then $\inf \sigma(H)$ is an eigenvalue.
\end{theoreme}
\begin{remark}
The condition $\delta>\frac{3}{2}$ in Thm. \ref{mainmain} comes from the operator bound $\omega^{-3}\leq C \langle x \rangle^{3+\epsilon}$, $\forall \ \epsilon>0$ proved in Thm. \ref{urgh}.
\end{remark}
\begin{remark}
From Lemma \ref{akito} we know that $\inf \sigma(\omega)=0$ if $\lim_{x\to \infty}m(x)=0$.
 Therefore the Nelson Hamiltonian can be massless using the terminology of Subsect. \ref{sec2.2}. 
\end{remark}
\begin{remark}
 In a subsequent paper \cite{GHPS2} we will show that if
 \[
0\leq m(x)\leq C\x^{-1-\epsilon}, \hbox{ for some }\epsilon>0,
\]
then $H$ has no ground state. Therefore the result of  Thm. \ref{mainmain} is sharp with respect to the decay rate of the mass at infinity.
\end{remark}
\subsection{Abstract Pauli-Fierz Hamiltonians}\label{sec4.1}
In \cite{BD}, Bruneau   and Derezi\'nski study
the spectral theory of abstract Pauli-Fierz Hamiltonians of the form
\[
H= K\otimes \one + \one\otimes \d\Gamma(\omega)+ \phi(v),
\]
acting on  the Hilbert space $\cH= {\mathcal K}\otimes\Gamma_{\s}(\ch)$, where $\cK$ is the Hilbert space for the small system 
and $\ch$ the one-particle space for the bosonic field. 
 The Hamiltonian $H$ is called massive (resp. massless) if $\inf
\sigma(\omega)>0$ (resp. $\inf \sigma(\omega)=0$). Among other results
they prove the existence of a ground state for $H$ if $v$ is infrared
regular.

Although most of  their hypotheses are natural and essentially optimal,
 we cannot directly apply their abstract results to our situation. 
In fact they assume (see \cite[Assumption E]{BD}) that the one-particle space $\ch$ equals $L^{2}(\rr^{d}, \d k)$ 
and the one-particle energy $\omega$ is the multiplication operator by a function $\omega(k)$ 
which is positive,  with $\nabla \omega$ bounded, and $\lim_{k\to \infty}\omega(k)=+\infty$. 
This assumption on the one-particle energy is only needed to prove an
HVZ theorem  for massive  (or massless with an infrared cutoff) Pauli-Fierz
Hamiltonians.

In our case this assumption  could be deduced (modulo unitary
equivalence) from the spectral theory of $h$. For example it would suffices to know
that $h$  is unitarily equivalent to $-\Delta$. This last property  would follow 
from the absence of eigenvalues for $h$ and from the scattering theory 
for the pair $(h, -\Delta)$ and 
require additional decay properties of  the $[a^{ij}](x)$, $m(x)$ and of some of their derivatives.

We will replace it by more geometric assumptions 
on $\omega$ (see hypothesis (\ref{e4.2}) below), similar to those introduce in \cite{GP}, 
where abstract bosonic QFT Hamiltonians were considered. 
Since we do not aim for generality, our hypotheses on the coupling
operator $v$ are stronger than necessary, but lead to simpler proofs. 
Also most of the proofs will be only sketched.

Let $\ch, \cK$ two Hilbert spaces and set $\cH= \cK\otimes \Gamma_{\s}(\ch)$.

We fix selfadjoint operators $K\geq 0$ on $\cK$ and  $\omega\geq 0$ on $\ch$. We set
\[
\inf\sigma(\omega)=:m\geq 0.
\] 
If $m=0$ one has to assume additionally that ${\rm Ker}\omega=\{0\}$ (see Remark \ref{ilote} for some explanation of this fact).
\begin{remark}\label{ilote}
It ${\mathcal X}$ is a real Hilbert space and $\omega$ is a selfadjoint operator on ${\mathcal X}$, the condition ${\rm Ker}\omega=\{0\}$  is well known to be necessary to have a stable quantization of the abstract Klein-Gordon equation 
$\p_{t}^{2}\phi(t)+ \omega^{2} \phi(t)=0$ where $\phi(t): \rr \to {\mathcal X}$.

If  ${\rm Ker}\omega\neq \{0\}$  the phase space ${\mathcal Y}={\mathcal X}\oplus {\mathcal X}$ 
for the Klein-Gordon equation splits into the symplectic direct sum ${\mathcal Y}_{\rm reg}\oplus {\mathcal Y}_{\rm sing}$, for  ${\mathcal Y}_{\rm reg}={\rm Ker}\omega^{\perp}\oplus {\rm Ker}\omega^{\perp}$, ${\mathcal Y}_{\rm sing}= {\rm Ker}\omega\oplus {\rm Ker}\omega$, both symplectic spaces being invariant under the symplectic evolution associated to  the Klein-Gordon equation. On ${\mathcal Y}_{\rm reg}$ one can perform the stable quantization. On ${\mathcal Y}_{\rm sing}$,if for example ${\rm Ker}\omega$ is $d-$dimensional, the quantization  leads to the Hamiltonian $-\Delta$ on $L^{2}(\rr^{d})$.  Clearly any perturbation of the form $\phi(f)$ for $\one_{\{0\}}(\omega)f\neq 0$ will make the Hamiltonian unbounded from below.
\end{remark}

So we will always assume that
\begin{equation}
\label{e4.0mass}
\omega\geq 0, \ {\rm Ker}\omega=\{0\}.
\end{equation}
Let $H_{0}= \cK\otimes\one+ \one\otimes \d\Gamma(\omega)$. We fix also
a coupling operator $v$ such that:
\beq
v\in B(\cK, \cK\otimes\ch).
\label{e4.0}
\eeq
The quadratic form $\phi(v)= a(v)+ a^{*}(v)$ is well defined for
example on $\cK\otimes \Dom N^{\12}$. 
We will also assume that:
\begin{equation}
\label{e4.4}
\omega^{-\12}v(K+1)^{-\12}\hbox{ is compact}.
\end{equation}
\begin{proposition}[\cite{BD} Thm. 2.2]
 Assume (\ref{e4.0mass}), (\ref{e4.4}). 
 Then $H= H_{0}+ \phi(v)$  is well defined as a form sum and yields a bounded below selfadjoint operator with $\Dom |H|^{\12}= \Dom |H_{0}|^{\12}$.
\end{proposition}
The operator $H$ defined as above is called an abstract Pauli-Fierz Hamiltonian.

\subsection{Existence of a ground state  for cutoff Hamiltonians}\label{sec4.2}
We introduce as in \cite{BD} the infrared-cutoff objects
\[
v_{\sigma}= F(\omega\geq \sigma)v, \ H_{\sigma}= K\otimes \one + \one \otimes \d\Gamma(\omega)+ \phi(v_{\sigma}), \ \sigma>0,
\]
where $F(\lambda\geq \sigma)$ denotes  as usual 
a function of the form $\chi(\sigma^{-1}\lambda)$, where  
$\chi\in \cinf(\rr)$, $\chi(\lambda)\equiv 0$ for $\lambda\leq 1$, $\chi(\lambda)\equiv 1$ for $\lambda\geq 2$. 

An important step to prove that $H$ has a ground state is to prove
that $H_{\sigma}$ has a ground state. The usual trick is to consider 
\[
\tilde{H}_{\sigma}= K\otimes\one+ \one \otimes\d\Gamma(\omega_{\sigma})+ \phi(v_{\sigma}),
\]
where:
\[
\omega_{\sigma}:= F(\omega\leq \sigma)\sigma + (1- F(\omega\leq
\sigma))\omega= \omega+ (\sigma- \omega)F(\omega\leq \sigma).
\]
\def\r{{\bf r}}
Note that since $\omega_{\sigma}\geq \sigma>0$, $\tilde{H}_{\sigma}$
is a massive Pauli-Fierz Hamiltonian. Moreover it is well known (see eg \cite{G}, \cite{BD}) $H_{\sigma}$ 
has a ground state iff $\tilde{H}_{\sigma}$ does.  The fact that
$\tilde{H}_{\sigma}$ has a ground state follows from an estimate on
its essential spectrum (HVZ theorem). In \cite{BD} this is shown using
the condition that $\ch=L^{2}(\rr^{d}, \d k)$ and $\omega= \omega(k)$.
Here we will replace this condition by the following more abstract
condition, formulated using an additional selfadjoint operator  $\r$ on
$\ch$. Similar abstract conditions were  introduced in \cite{GP}.

\medskip

We will assume that there exists an  selfadjoint operator $\r\geq 1$ on $\ch$ such 
that the following conditions hold for all $\sigma>0$:
\begin{equation}
\label{e4.2}
\begin{array}{l}
(i) \ (z-\r)^{-1}:\ \Dom \omega_{\sigma}\to\Dom \omega_{\sigma}, \
\forall \  z\in \cc\backslash \rr ,\\[2mm]
(ii) \ [\r, \omega_{\sigma}] \hbox{ defined as a quadratic form on }\Dom \r\cap \Dom \omega\hbox{ is bounded}, \\[2mm]
(iii) \ \r^{-\epsilon}(\omega_{\sigma}+1)^{-\epsilon}\hbox{ is compact on
}\ch \hbox{ for some }0<\epsilon< \12.
\end{array}
\end{equation}
The  operator $\r$, called a {\em gauge}, is used  to localize particles in $\ch$. 

We assume also as in \cite{BD}:
\begin{equation}
\label{e4.5}
(K+1)^{-\12}\hbox{ is compact}.
\end{equation}
This assumption means that the small system is confined. 

\begin{proposition}\label{irut}
 Assume (\ref{e4.0mass}), (\ref{e4.0}), (\ref{e4.4}), (\ref{e4.2}),(\ref{e4.5}) . Then
\[
\sigma_{\rm ess}(\tilde{H}_{\sigma})\subset [\inf
\sigma(\tilde{H}_{\sigma})+ \sigma, +\infty[.
\]
It follows that $\tilde{H}_{\sigma}$ (and hence $H_{\sigma}$) has a ground state for all $\sigma>0$.
\end{proposition}
\proof By (\ref{e4.4}), $\phi(v_{\sigma})$ is form bounded with
respect  to $H_{0}$ (and to  $K\otimes\one +
\one\otimes\d\Gamma(\omega_{\sigma})$) with the
infinitesimal bound, hence $H_{\sigma}, \ \tilde{H}_{\sigma}$ are well
defined as  bounded below selfadjoint Hamiltonians. 

 We can follow the proof of \cite[Thm. 4.1]{DG2} or \cite[Thm.
7.1]{GP} for its abstract version.  For ease of  notation we denote
simply $\tilde{H}_{\sigma}$ by $H$, $\omega_{\sigma}$ by $\omega$ and
$v_{\sigma}$ by $v$.
The key estimate is the fact that for $\chi\in \coinf(\rr)$ one has
\begin{equation}
\label{e4.7}
\chi(H^{\rm ext}) I^{*}(j^{R}) - I^{*}(j^{R})\chi(H)\in o(1), \hbox{
when }R\to \infty.
\end{equation}
(The extended operator $H^{\rm ext}$ and identification operator $I(j^{R})$ are defined for example in \cite[Sect.2.4]{GP}). 
The two main ingredients of the proof of (\ref{e4.7}) are the estimates:
\begin{equation}
\label{e4.8}
[F(\frac{\r}{R}), \omega_{\sigma}]\in O(R^{-1}), \ F\in \coinf(\rr),
\end{equation}
and
\begin{equation}
\label{e4.9}
\omega_{\sigma}^{-\12}F(\frac{\r}{R}\geq 1) v_{\sigma}(K+1)^{-\12}\in o(R^{0}).
\end{equation}
Now (\ref{e4.9}) follows  from the fact that $v_{\sigma}(K+1)^{-\12}$ is
compact (note that $\omega_{\sigma}^{-\12}$ is bounded since
$\omega_{\sigma}\geq \sigma$),  and  (\ref{e4.8}) follows  from  Lemma
\ref{irta}.
The estimate (\ref{e4.7}) can  then be proved exactly as in \cite[Lemma 6.3]{GP}.
 Note that here we prove only the $\subset$ part of the HVZ theorem, which is sufficient for our purposes. 
The details are left to the reader. \qed

\begin{lemma}\label{irta}
 Assume conditions (i), (ii) of (\ref{e4.2}). Then for all $F\in \coinf(\rr)$ one has:
 \[
\begin{array}{l}
F(\r): \ \Dom \omega_{\sigma}\to \Dom \omega_{\sigma}, \\[2mm]
[F(\frac{\r}{R}), \omega_{\sigma}]\in O(R^{-1}).
\end{array}
\]
\end{lemma}
\proof
The proof of the lemma is easy, using almost analytic extensions, as
for example in \cite{GP}. The details are left to the interested
reader. \qed

\subsection{Existence of a ground state for massless models}\label{sec4.3}
Let us  introduce the following hypothesis on the coupling operator
(\cite[Hyp. F]{BD}):
\beq
\omega^{-1}v(K+1)^{-\12}\hbox{ is compact}.
\label{e4.10}
\eeq
\begin{theoreme}\label{iriat}
Assume (\ref{e4.0mass}), (\ref{e4.0}), (\ref{e4.4}),  (\ref{e4.2}), (\ref{e4.5}) and
(\ref{e4.10}). Then $H$ has a ground state.
\end{theoreme}
\proof we can follow the proof in \cite[Sect. 4]{BD}. The existence of
ground state for $H_{\sigma}$ (\cite[Prop. 4.5]{BD}) is shown in
Prop. \ref{irut}.  The arguments in \cite[Sects 4.2, 4.3]{BD} based on
the pull-through and double pull-through formulas are abstract and valid
for any one particle operator $\omega$. The only place where the fact
that $\ch= L^{2}(\rr^{d}, \d k)$ and $\omega= \omega(k)$ appears is in \cite[Prop. 4.7]{BD} where the operator $|x|= |\i \nabla_{k}|$
enters. In our situation it suffices to replace it by our gauge
operator $\r$. The rest of the proof is unchanged. \qed
\subsection{Proof of Thm. \ref{mainmain}}
We  now complete the proof of Thm. \ref{mainmain}, by verifying the
hypotheses of Thm. \ref{iriat}. We recall that $\ch= L^{2}(\rr^{d} \d
x)$, $\omega= h^{\12}$  and
we will take $\r=\x=(1+x^{2})^{\12}$.
\medskip

{\bf Proof of Thm. \ref{mainmain}.}

 We saw in the proof of Lemma \ref{0.1} that $v$, $\omega^{-\12}v$ are bounded, hence in particular (\ref{e4.0}) is satisfied. 
 By hypothesis (E2),  $(K+1)^{-\12}$ is compact, 
 which implies  that conditions  (\ref{e4.4}) and (\ref{e4.5}) are satisfied. 
 
We now check condition (\ref{e4.2}). 
Note that  $\omega_{\sigma}= f(h)$
where $f\in \cinf(\rr)$ with $f(\lambda)= \lambda^{\12}$ for
$\lambda\geq 2$. Clearly $\Dom \omega_{\sigma}= H^{1}(\rr^{d})$ which
is preserved by $(z-\x)^{-1}$, so (i) of (\ref{e4.2}) is satisfied.
Condition (iii) is also obviously satisfied.  It remains to check
condition (ii).  To this end we write $\omega_{\sigma}= f(h)=
(h+1)g(h)$ where $g\in \cinf(\rr)$ satisfies
\[
g^{(n)}(\lambda)\in O(\langle \lambda\rangle^{-\12-n}), \
n\in \nn,
\]
and hence 
\beq\label{e4.11}
[\x, \omega_{\sigma}]= [\x, h]g(h) + (h+1) [\x, g(h)].
\eeq
Since $\nabla a^{jk}(x)$, $\nabla c(x)$, $\nabla m(x)$ are bounded and $\Dom h=H^{2}(\rr^{d})$ we
see that
\beq\label{e4.11bis}
[\x, h](h+1)^{-\12}, [[\x, h],h](h+1)^{-1}\hbox{ are  bounded}.
\eeq
In particular the first term in the
r.h.s. of (\ref{e4.11}) is bounded. To estimate the second term, we
use an almost analytic extension of $g$ satisfying:
\beq\label{e4.10bis}
\begin{array}{l}
\tilde{g}_{|\rr}= g, \ |\frac{\p \tilde{g}}{\partial
\overline{z}}(z)|\leq C_{N}\langle z\rangle^{-3/2-N}|{\rm Im}z|^{N}, \
N\in \nn, \\[2mm]
{\rm supp} \tilde{g}\subset \{z\in \cc| |{\rm Im}z|\leq c(1+ |{\rm
Re}z|)\},
\end{array}
\eeq
(see eg \cite[Prop. C.2.2]{DG}), and write
\[
g(h)=  \frac{\i}{2\pi}\int_{\cc}\frac{\p{\tilde{g}}}{\p 
\zbar}(z)(z-h)^{-1}\d z\wedge \d\zbar.
\]
We perform a  commutator expansion to obtain that:
\[
[\x, g(h)]= g'(h)[\x, h]+ R_{2},
\]
for
\[
R_{2}= \frac{\i}{2\pi}\int_{\cc}\frac{\p{\tilde{g}}}{\p 
\zbar}(z)(z-h)^{-2}[[\x, h]h](z-h)^{-1}\d z\wedge \d\zbar.
\]
Since $|g'(\lambda)|\leq C\langle \lambda\rangle^{-3/2}$,
$(h+1)g'(h)[\x, h]$ is bounded. To estimate the term $(h+1)R_{2}$, we
use  again (\ref{e4.11bis}) and the bound
\[
\|(h+1)^{\alpha}(z-h)^{-1}\|\leq C \langle z\rangle^{\alpha}|{\rm
Im}z|^{-1}, \ \alpha= \12, 1.
\]
We obtain that
\[
\|(h+1)R_{2}\|\leq C \|[[\x, h]h](h+1)^{-1}\|\int_{\cc}|\frac{\p{\tilde{g}}}{\p 
\zbar}(z)| \langle z\rangle^{2}|{\rm Im}z|^{-3}\d z\d\zbar.
\]
This integral is convergent using the estimate (\ref{e4.10bis}). This
completes the proof of (\ref{e4.2}).  

It remains to check condition (\ref{e4.10}), i.e. the fact that the interaction is infrared regular. This is the only place where the lower bound on $m(x)$ enters. By  Thm. \ref{urgh} we obtain that $\omega^{-3/2}\x^{-3/2-\epsilon}$ is bounded for all  $\epsilon>0$.  By condition (E2), we obtain that $\langle X\rangle^{3/2 + \epsilon}(K+1)^{-\12}$ is bounded for all  $\epsilon>0$ small enough.

Therefore to check (\ref{e4.10}) it suffices to prove that the map
\[
L^{2}(\rr^{3}, \d X)\ni u\mapsto \x^{3/2+\epsilon}\rho(x-X)\langle X\rangle^{-3/2-\epsilon}u(X)\in L^{2}(\rr^{3}, \d X)\otimes L^{2}(\rr^{3}, \d x)
\]
is bounded, which is immediate since $\rho\in S(\rr^{3})$. This completes the proof of Thm. \ref{mainmain}.
\qed
\subsection{Existence of a ground state for non confined Hamiltonians}\label{non-confining}
In this subsection we state the results on existence of a ground state if the electronic potential is not confining. As explained in the beginning of this section, one has to assume a {\em stability condition}, meaning that states near the bottom of the spectrum of $H$ are confined in electronic variables from energy conservation arguments.  
\begin{definition}
 Let $H$ be a Nelson Hamiltonian  satisfying (E1), (B1). We assume  for simplicity that the  electronic potential $W(X)$ is bounded. Set for $R\geq 1$:
 \[
D_{R}= \{u\in \Dom H\ | \one_{\{|X|\leq R\}}u=0\}.
\]
The {\em ionization threshold} of $H$ is
\[
\Sigma(H):=\lim_{R\to +\infty}\inf_{u\in D_{R}, \ \|u\|=1}(u|Hu).
\]
\end{definition}
The following theorem can easily be obtained by adapting the arguments in this section.
\begin{theoreme}
 Assume hypotheses (E1), (B1), $W\in L^{\infty}(\rr^{3})$ and  $m(x)\geq a \langle x\rangle^{-1}$ for some $a>0$.  Then if the following {\em stability condition} is satisfied:
 \[
\Sigma(H)>\inf \sigma(H),
\]
$H$ has a ground state.
\end{theoreme}
{\bf Sketch of proof.} Assuming the stability condition one can prove using Agmon-type estimates as in \cite{Gr}  (see \cite{P} for the case of the Nelson model)
that if $\chi\in \coinf(]-\infty, \Sigma(H)[$ then $\e^{\beta|X|}\chi(H_{\sigma})$ is bounded uniformly in 
$0<\sigma\leq \sigma_{0}$ for $\sigma_{0}$ small enough. 
From this  fact one deduces by the usual argument that $H_{\sigma}$ has a ground state $\psi_{\sigma}$ and that  
\begin{equation}
\label{uargh}
\sup_{\sigma>0}\|\langle X\rangle^{N}\psi_{\sigma}\|<\infty.
\end{equation}  One can then follow the proof in  \cite[Thm. 1.2]{P}. The key infrared regularity property replacing (\ref{e4.10}) is now
\[
\sup_{\sigma>0}\|\omega^{-1}v \psi_{\sigma}\|_{\cH\otimes \ch}<\infty.
\]
This estimate follows as in the proof of (\ref{e4.10}) from Thm. \ref{urgh} and the bound (\ref{uargh}). The details are left to the reader. \qed

\appendix
\section{Lower bounds for second order differential
operators}\init\label{sec1}
In this section we prove various lower bounds for second order
differential operators. These bounds are the
key ingredient in the proof of the existence of a ground state for the Nelson model.
\subsection{Second order differential operators}\label{sec1.1}
Let us introduce the class of second order differential operators that will be studied in this section.
Let:
\[
\begin{array}{rl}
h_{0}=&\sum_{1\leq j,k\leq d} c(x)^{-1}D_{j}a^{jk}(x)D_{k}c(x)^{-1},\\[2mm]
 h=& h_{0}+ v(x),
\end{array}
\]
with $a^{jk}$, $c$, $v$ real functions and:
\beq
\begin{array}{l}
C_{0}\one\leq [a^{jk}(x)]\leq C_{1}\one,  \ C_{0}\leq c(x)\leq C_{1}, \ C_{0}>0, \\[2mm]
\p_{x}^{\alpha}a^{jk}(x)\in O(\langle x\rangle^{-1}), \ |\alpha|\leq 1, \ \p_{x}^{\alpha}c(x)\in O(1), \ |\alpha|\leq 2,
\end{array}
\label{e1.1}
\eeq
\beq
v\in L^{\infty}(\rr^{d}), \ v\geq 0.
\label{e1.1bis}
\eeq
Clearly $h_{0}$  and $h$ are  selfadjoint and positive with domain $H^{2}(\rr^{d})$. We will always assume that $d\geq 3$.
\subsection{Upper bounds on heat kernels}\label{sec1.2}

If $K$ is a bounded op\-er\-at\-or on $L^{2}(\rr^{d}, c^{2}\d x)$ we will denote by $K(x, y)\in {\mathcal D'}(\rr^{2d})$ its distribution kernel.
In this subsection we will prove the following theorem.
We set:
\[
\psi_\alpha(t,x) := 
\left(\frac{\x^{2}}{\x^{2}+t}\right)^{\alpha}, \ \alpha>0.
\]
\begin{theoreme} \label{main}
Assume in addition to (\ref{e1.1}), (\ref{e1.1bis}) that:
\[
v(x)\geq a\x^{-2}, \ a>0,
\]
then  there exists $C, c,\alpha>0$ such that:
\beq\label{oro12}
 e^{-th}(x,y) \leq  C \psi_\alpha(t,x)\psi_\alpha(t,y) e^{c t
\Delta}(x,y), \ \forall\ t>0, \ x,y\in \rr^{d}. 
\eeq
\end{theoreme}
If $c(x)\equiv 1$ or if $h_{0}$ is the Laplace-Beltrami operator for a Riemannian metric on $\rr^{d}$, then Thm. \ref{main} is due to Zhang \cite{Zh}.
\begin{remark}\label{ilt}
Conjugating by the unitary
\[
U: \ \begin{array}{rl}
L^{2}(\rr^{d}, \ \d x)\to &L^{2}(\rr^{d}, c^{2}(x)\d x),\\[2mm]
u\mapsto& c(x)^{-1}u,
\end{array}
\]
we obtain
\[
\begin{array}{rl}
\tilde{h}_{0}:=&Uh_{0}U^{-1}= c(x)^{-2}\sum_{1\leq j,k\leq d} D_{j}a^{jk}(x)D_{k}, \\[2mm]
  \tilde{h}:=&UhU^{-1}= \tilde{h}_{0}+ v(x),
\end{array}
\]
which are selfadjoint with domain $H^{2}(\rr^{d})$. 
 Let $\e^{-t\tilde{h}}(x, y)$  for $t>0$ the integral kernel of $\e^{-t\tilde{h}}$ i.e.  such that
 \[
\e^{-t\tilde{h}}u(x)=\int_{\rr^{d}}\e^{-th}(x, y)u(y)c^{2}(y)\d y, \ t>0.
\]
Then since $\e^{-th}(x, y)= c(x)\e^{-t\tilde{h}}(x, y)c(y)$, it suffices to prove Thm. \ref{main} for $\e^{-t\tilde{h}}$. 
\end{remark}
By the above remark, we will consider the operator $\tilde{h}_{0}$ (resp. $\tilde{h}$) and denote it again by $h_{0}$ (resp. $h$).
We note that they are associated with the closed quadratic forms:
\[
Q_{0}(f)=\int_{\rr^{d}}\sum_{j,k}\p_{j}\overline{f}a^{jk}\p_{k}f\ \d x, \\ Q(f)= Q_{0}(f)+ \int_{\rr^{d}} |f|^{2}c^{2}v \ \d x,
\] 
with domain $H^{1}(\rr^{d})$.

Let us consider the semi-group $\{\e^{-th}\}_{t\geq 0}$ generated by $h$.  Since $\Dom Q_{0}= H^{1}(\rr^{d})$, we can apply \cite[Thms. 1.3.2, 1.3.3]{Da}
to obtain that 
 $\e^{-th}$ is positivity preserving and  extends as a semi-group of contractions on $L^{p}(\rr^{d}, c^{2}\d x)$ for $1\leq d\leq \infty$, strongly continuous on $L^{p}(\rr^{d}, c^{2}\d x)$ if $p<\infty$.   In other words $\{\e^{-th}\}_{t\geq 0}$  is a Markov symmetric semigroup.
  
  We first recall two results, taken from \cite{PE} and \cite{Da}.
\begin{lemma}\label{blop}
 Assume  (\ref{e1.1}), (\ref{e1.1bis}). Then there exist $c, C>0$ such that:
 \[
0\leq \e^{-th}(x, y)\leq C\e^{ct\Delta}(x, y), \forall \ 0<t, \ x, y\in \rr^{d}.
\]
\end{lemma}
\proof Since $v(x)\geq 0$ it follows from  the Trotter-Kato formula that
\[
0\leq \e^{-th}(x, y)\leq \e^{-th_{0}}(x, y), \hbox{ a.e. }x, y.
\]
The stated upper bound on $\e^{-th_{0}}(x, y)$ is shown in \cite[Thm. 3.4]{PE}. \qed

The following lemma is an extension of  \cite[Lemma 2.1.2]{Da} where the case $c(x)\equiv 1$ is considered.
 \begin{lemma}\label{limi}
Assume (\ref{e1.1}), (\ref{e1.1bis}). Then:
\ben
\item $e^{-th}$ is ultracontractive, i.e.
$e^{-th}$ is bounded from $L^2$ to $L^\infty$ for all $t>0$,
and 
\[ c_t := \|e^{-th}\|_{L^2 \to L^\infty} = \sup_{f \in L^2} 
	\frac{\|e^{-th}f\|_\infty}{\|f\|_2} \leq c t^{-d/4} \]
with some constant $c >0$.
\item $e^{-th}$ is bounded  from $L^1$ to $L^\infty$ 
for all $t>0$ and
\[ \|e^{-th}\|_{L^1 \to L^\infty} \leq c_{t/2}^2. \]
\item The kernel $e^{-th}(x,y)$ satisfies:
\[ 0 \leq e^{-th}(x,y) \leq c_{t/2}^2. 
\]
\een
\end{lemma}
\proof
  From Lemma \ref{blop} we obtain that 
  \[
\|\e^{- th} f\|_{\infty}\leq C\|\e^{ct\Delta}|f|\|_{\infty}\leq C't^{- d/4}\|f\|_{2},
\]
using the explicit form of the heat kernel of the Laplacian. This proves (1).

Taking adjoints  we see that $\e^{-th}$ is also bounded  from $L^{1}$ to $L^{2}$ with\\ $\|e^{-th}\|_{L^1 \to L^2} \leq c_{t}$. It follows that
\[
\|\e^{-th}\|_{L^{1}\to L^{\infty}}\leq\|\e^{-th/2}\|_{L^{2}\to L^{\infty}}\|\e^{-th/2}\|_{L^{1}\to L^{2}}\leq c_{t/2}^{2},
\]
which proves (2). Statement (3) is shown in \cite[Lemma 2.1.2]{Da}. \qed

\medskip

We will deduce Thm. \ref{main} from the following result.

\begin{theoreme} \label{heatkernelbound}
Assume the hypotheses of Thm. \ref{main}. Then there exists $C,\alpha>0$ such that:
\[ 
e^{-th}(x,y) \leq  Ct^{-d/2} \psi_\alpha(t,x)\psi_\alpha(t,y).
\]
\end{theoreme}
{\bf Proof of Theorem \ref{main}:}

Combining Lemma \ref{blop} with Thm. \ref{heatkernelbound} we get:
\[
\begin{array}{rl}
\e^{-th}(x, y)=&\left(\e^{-th}(x, y)\right)^{\epsilon}\left(\e^{-th}(x, y)\right)^{1-\epsilon}\\[2mm]
\leq &Ct^{-\epsilon d/2}\e^{-\epsilon(x-y)^{2}/2t}t^{-(1-\epsilon)d/2}\psi_\alpha(t,x) ^{1-\epsilon}\psi_\alpha(t,y)^{1-\epsilon}\\[2mm]
\leq & C't^{-d/2}\e^{-c(x-y)^{2}/2t} \psi_{\beta}(t, x)\psi_{\beta}(t, y),
\end{array}
\]
for $\beta=(1-\epsilon)\alpha$. This completes the proof of Thm. \ref{main}. \qed

It remains to prove Theorem \ref{heatkernelbound}.
To this end, we employ the following abstract result.
\begin{lemma} \label{Simon}(\cite[Theorem B]{MS}) 
Let $(M,d\mu)$ be a locally compact measurable space with
$\sigma$-finite measure $\mu$ and let $A$ be a non-negative self-adjoint operator
on $L^2(M,d\mu)$ such that
\begin{enumerate}
\item[(i)] 
$e^{-t A_1}:=(e^{-tA}|_{L^1\cap L^2})^{\rm clos}_{L^1 \to L^1}$,
$t \geq 0$ is a $C_0$-semi-group of bounded operators, i.e.,
\[ \|e^{-t A_1}\|_{L^1 \to L^1} \leq c_1,
	\quad t \geq 0. \]
\item[(ii)] $e^{-tA}$ is bounded from $L^{1}$ to $L^{\infty}$ with:
\[ \|e^{-tA_1}\|_{L^1 \to L^\infty} \leq c_2 t^{-j},
	\quad t > 0,
	 \]
for some $j > 1$.
\end{enumerate}
Assume moreover that there exists a family of weights 
$\psi(s,x)$ ($s>0$) such that: 
\begin{enumerate}
\item[($B1$)] 
$\psi(s,x)$, $\psi(s,x)^{-1} \in L^2_{\rm loc}(M\setminus N,d\mu)$
for all $s>0$, where $N$ is a closed null set.
\item[($B2$)]
There is a constant $\tilde{c}$ independent of $s$
such that, for all $t \leq s$,
\[ \|\psi(s,\cdot)e^{-tA}\psi(s,\cdot)^{-1}f\|_1 \leq \tilde{c}\|f\|_1,
	\quad f \in D_s, \]
where $D_s := \psi(s,\cdot)L^\infty_{\rm c}(M \setminus N, d\mu)$
\item[($B3$)]
There exists $0<\epsilon<1 $ and constants $\hat{c}_{i}>0$, $i=1,2$ such that for any  $s>0$
there exists a measurable set $\Omega^s \subset M$ with\item[(a)]
$|\psi(s,x)|^{-\epsilon} \leq \hat{c}_1$ for all $x \in M \setminus \Omega^s$,
\item[(b)]
$|\psi(s,x)|^{-\epsilon} \in L^{q}(\Omega^s)$ and
$\||\psi(s,\cdot)|^{-\epsilon}\|_{L^{q}(\Omega^s)} \leq \hat{c}_2s^{j/q}$
with $q=2/(1-\epsilon)$ and $j>1$ is the exponent in condition (ii).
\end{enumerate}
Then there is a constant $C$ such that 
\[
 |e^{-tA}(x,y)| \leq Ct^{-j}|\psi(t,x)\psi(t,y)|, \ \forall \  t>0, \ \hbox{a.e. }x, y \in M. 
 \]
\end{lemma}

To verify condition (B2) of Lemma \ref{Simon}, we will  use the following lemma. 
\begin{lemma}\label{criter} (\cite[Criterion 2]{MS})
 Let $e^{-tA}$ be a $C_0$-semi-group on $L^2(M,d\mu)$. Denote by $\langle\cdot, \cdot \rangle$ 
the scalar product on $L^{2}(M, \d \mu)$.
Then:
\[
 \| e^{-tA}f\|_{L^\infty} \leq \|f\|_{L^\infty},
\quad f \in L^2 \cap L^\infty, \quad t>0,
 \]
if and only if:
\beq\label{criterion} {\rm Re} \langle f-f_{\wedge}, Af \rangle \geq 0, 
\quad f \in D(A), \eeq
where $f_{\wedge} = (|f| \wedge 1){\rm sgn} f$ 
with ${\rm sgn}f(x) :=f(x)/|f|(x)$ if $|f|(x)\not=0$ and ${\rm sgn}f(x)=0$ if $f(x)=0$.\end{lemma}

{\bf Proof of Thm. \ref{heatkernelbound}:} 
We will prove that there exists $\alpha>0$ such that the hypotheses of Lemma \ref{Simon} are satisfied for
$(M, \d \mu)=(\rr^{d}, c^{2}(x)\ d x)$, $A= h$ and $\psi(s, x)= \psi_{\alpha}(s, x)$. For ease of notation 
we will  often denote $\psi_{\alpha}$ simply by $\psi$.

From the discussion before Lemma \ref{limi}, we know that $\e^{-th}$ extends as a $C_{0}-$semi-group of contractions of $L^{1}(\rr^{d}, c^{2}\d x)$, which implies that  hypothesis (i) holds with $c_1=1$.  Hypothesis (ii) with $j=d/2$ follows from (2) of Lemma \ref{limi}. Note that $d/2>1$ since $d\geq 3$.

We now check that conditions (B) are satisfied by $\psi_{\alpha}$
provided we choose $\alpha= \alpha_{0}a^{\12}$ for some constant
$\alpha_{0}$. Since $\psi, \ \psi^{-1}$ are bounded, condition (B1) is satisfied for all $\alpha>0$.  Set $\Omega^s := \{ x\in \mathbb{R}^d \mid \langle x \rangle^2 \leq s \}$.
Then 
\[ \psi(x)^{-\epsilon} 
= \left[
	\frac{\langle x\rangle^2 + s}{\langle x\rangle^2}\right]^{\alpha\epsilon}
\leq 2^{\alpha\epsilon},\ \forall \ x\not\in \Omega^{s},
 \]
which proves  the bound (a)  of (B3) for all $\alpha>0$.
Take now  $0<\epsilon < \frac{d}{d+4\alpha}$ so that 
we see that $d - 2\alpha\epsilon q > 0$ for $q=2/(1-\epsilon)$. If $0\leq s<1$ $\Omega^{s}=\emptyset$ and (b) of (B3) is satisfied. If $s\geq 1$ we have:
\begin{align*}
\|\psi^{-\epsilon}\|_{L^q(\Omega^s)}^q
& = \int_{\Omega^s} \left[ 
	\frac{\langle x\rangle^2 + s}{\langle x\rangle^2}
	\right]^{\alpha\epsilon q} c^{2}(x)\d x \\
& \leq C_{1}^{2}(2s)^{\alpha\epsilon q}
	\int_{\{|x| \leq \sqrt{s}\}} |x|^{-2\alpha\epsilon q}\d x \\
& = Cs^{\alpha\epsilon q}
	\int_{0}^{\sqrt{s}}  r^{d-2\alpha\epsilon q-1}\d r  = C' s^{d/2}. 
\end{align*} 
Hence (b) is satisfied for  $j=d/2$.

It remains to check (B2). 
To avoid confusion, we denote by $\langle g, f\rangle$ the scalar product in $L^{2}(\rr^{d}, c^{2}(x)\d x)$ and by $(g|f)$ the usual scalar product in 
$L^{2}(\rr^{d}, \d x)$.

Since $\psi, \ \psi^{-1}$ are $C^{\infty}$ and bounded with all derivatives, we see that\\ $ \{\psi\e^{-th}\psi^{-1}\}_{t\geq 0}$ is a $C_{0}-$semi-group on $L^{2}(\rr^{d}, c^{2}\d x)$, with generator
\[
h_{\psi}:= \psi h \psi^{-1}, \ \Dom h_{\psi}= H^{2}(\rr^{d}).
\]
We claim that there exists $\alpha>0$ such that 
\begin{equation}
\label{tutu}
\|\e^{-th_{\psi}}\|_{L^{1}\to L^{1}}\leq C, \hbox{ uniformly for }0\leq t\leq s.
\end{equation}
By duality, (\ref{tutu}) will follow from (\ref{tututu}):
\begin{equation}
\label{tututu}
\|\e^{-th^{*}_{\psi}}\|_{L^{\infty}\to L^{\infty}}\leq C, \hbox{ uniformly for }0\leq t\leq s.
\end{equation}
To prove (\ref{tututu}), we will apply Lemma \ref{criter}.  To avoid confusion, $\p_{j}f(x)$ will denote a partial derivative of  the function $f$ , 
while $\nabla_{j}f(x)$  denote the product of the operator $\nabla_{j}$ and the operator of multiplication by  the function $f$.

Setting $b_{i}=\psi^{-1}\p_{i}\psi$, we have:
\[
\begin{array}{rl}
h^{*}_{\psi}= &\psi^{-1}h\psi\\[2mm]
=&-c(x)^{-2}\sum_{ j,  k}\nabla_{j}a^{jk}(x)\nabla_{k} - \sum_{ j,  k}c^{-2}(x)b_{j}(x)a^{jk}(x)\nabla_{k}\\[2mm] -&c^{-2}(x)\nabla_{j} a^{jk}(x)b_{k}(x)
+ v(x)-c^{-2}(x)\sum_{ j,  k}b_{j}(x)a^{jk}(x)b_{k}(x)\\[2mm]
=&-c(x)^{-2}\sum_{ j,  k}\nabla_{j}a^{jk}(x)\nabla_{k} - 2c(x)^{-2}\sum_{ j,  k}b_{j}(x)a^{jk}(x)\nabla_{k} + w(x),
\end{array}
\]
\def\sgn{{\rm sgn}}
where:
\[
\begin{array}{rl}
w(x)=& v(x)-c(x)^{-2}\sum_{ j,  k}b_{j}(x)a^{jk}(x)b_{k}(x)\\[2mm]
-& c(x)^{-2}\sum_{j,k}a^{jk}(x)\p_{j}b_{k}(x)-c(x)^{-2}\sum_{j,k}(\p_{j}a^{jk})(x)b_{k}(x).
\end{array}
\]

%

Clearly $\Dom h_{\psi}^{*}= H^{2}(\rr^{d})$.  
To simplify notation, we set 
$A(x)= [a^{jk}(x)]$,   $F(x)= (b_{1}(x), \dots, b_{d}(x))$. The identity above becomes:
\beq\label{oro6}
\begin{array}{rl}
h_{\psi}^{*}=& -c^{-2}\nabla_{x}A \nabla_{x} - c^{-2}F A\nabla_{x}- c^{-2}\nabla_{x} AF+ v- c^{-2}FAF,\\[2mm]
=&  -c^{-2}\nabla_{x}A \nabla_{x} - 2c^{-2}F A\nabla_{x} + w.
\end{array}
\eeq
We note that $b_{j}(x)= \alpha sx_{j}\x^{-2}(\x^{2}+ s)^{-1}$, which
implies that:
\[
|b_{j}(x)|\leq C\alpha \x^{-1}, \ |\nabla_{x}b_{j}(x)|\leq C \alpha
\x^{-2}, \hbox{for some }C>0.
\]
Since $v(x)\geq a\x^{-2}$,  this implies using also (\ref{e1.1}) that:
\beq
v(x)-c(x)^{-2}FAF(x)\geq 0, \ w(x)\geq 0, 
\label{oro11}
\eeq 
for $\alpha>0$ small enough. 


%
This implies that
\beq\label{oro7}
{\rm Re} \langle f, h_{\psi}^{*}f\rangle =  -(\nabla_{x}f|A \nabla_{x}f)+ (f|(c^{2}v- FAF)f)\geq 0, \ \hbox{ for }f\in H^{1}(\rr^{d}).
\eeq
 It follows that $h_{\psi}^{*}$ is maximal accretive, hence $\e^{- th_{\psi}^{*}}$ is a $C_{0}-$semi-group of contractions
by the Hille-Yosida theorem.

To check condition (\ref{criterion}) in Lemma \ref{criter} we follow \cite{MS}, with some easy modifications. We write
\[
f-f_{\Lambda}= \sgn f \chi, \ \chi:=\one_{\{|f|\geq 1\}}(|f|-1),
\]
and note that if $f\in \Dom  h_{\psi}^{*}\subset H^{1}(\rr^{d})$ then $|f|, \ \sgn f, \ \chi\in H^{1}(\rr^{d})$ with
\beq\label{irit}
\nabla \sgn f= \frac{\nabla f}{|f|}- f\frac{\nabla f}{|f|^{2}}, \ \nabla\chi= \one_{\{|f|\geq 1\}}\nabla |f|, \ \nabla |f|= \frac{1}{2|f|}(\overline{f}\nabla f+ f\nabla\overline{f} ).
\eeq
 We have:
\[
\begin{array}{rl}
\langle f-f_{\Lambda},\ h_{\psi}^{*}f\rangle=&(\nabla (f- f_{\Lambda})| A \nabla f)-2(F (f- f_{\Lambda})| A\nabla f)+ ( (f- f_{\Lambda})| c^{2}w f) \\[2mm]
=:&C_{1}(f)+ C_{2}(f)+ C_{3}(f).
\end{array}
\]
Using (\ref{irit}), we have:
\[
\begin{array}{rl}
C_{1}(f)=& (\nabla (f- f_{\Lambda})| A \nabla f)\\[2mm]
=& (\nabla f|\frac{\chi}{|f|}A\nabla f)-(\nabla |f| |\overline{f} \frac{\chi}{|f|^{2}}A\nabla f)+ (\nabla \chi|\frac{\overline{f}}{|f|}A\nabla f)\\[2mm]
=:& B_{1}(f)+ B_{2}(f)+ B_{3}(f).
\end{array}
\]
Clearly $B_{1}(f)$ is real valued. Next:
\beq\label{oro1}
{\rm Re} B_{2}(f)=-\12 (\nabla |f|| \frac{\chi}{|f|^{2}}A (\overline{f}\nabla f+ f\nabla\overline{f}))
= -(\nabla |f|| \frac{\chi}{|f|}A \nabla |f|),
\eeq
using (\ref{irit}). Similarly:
\beq\label{oro2}
{\rm Re}B_{3}(f)=\12(\nabla \chi|\frac{1}{|f|} A(\overline{f}\nabla f+ f\nabla\overline{f}))=(\nabla \chi| A\nabla \chi),
\eeq
using again (\ref{irit}). 
We estimate now ${\rm Re}C_{2}(f)$. We have:
\beq\label{oro4}
{\rm Re} C_{2}(f)=-2{\rm Re}( F(f-f_{\Lambda})| A\nabla f)=\12(\chi|\frac{F}{|f|}A(\overline{f}\nabla f+ f\nabla\overline{f}))=-2(F\chi| A\nabla \chi).
\eeq
We estimate now ${\rm Re} C_{3}(f)$. We have:
\beq\label{oro3}
{\rm Re} C_{3}(f)={\rm Re}( f-f_{\Lambda}|c^{2}w f)= {\rm Re}( \chi| c^{2}w |f|)=(\chi|c^{2}w|f|)= (\chi|c^{2} w \chi)+ ( \chi| c^{2}w).
\eeq
Collecting (\ref{oro1}) to (\ref{oro4}), we obtain that:
\begin{equation}
\label{oro5}
\begin{array}{rl}
{\rm Re}\langle f-f_{\Lambda}, h_{\psi}^{*}f\rangle=& (\nabla f|\frac{\chi}{|f|}A\nabla f)-(\nabla |f|| \frac{\chi}{|f|}A \nabla |f|)\\[2mm]
&+(\nabla \chi| A\nabla \chi)-2(F\chi| A\nabla \chi)+(\chi| c^{2}w \chi).\\[2mm]
&+ ( \chi| c^{2}w).
\end{array}
\end{equation}
We use now the point-wise identity:
\[
\begin{array}{rl}
&\nabla\overline{f}A \nabla f-\nabla|f|A\nabla |f|\\[2mm]
=& \nabla\overline{f}A \nabla f-\frac{1}{4|f|^{2}}(\overline{f}\nabla f+ f\nabla |f|)A(\overline{f}\nabla f+ f\nabla |f|)\\[2mm]
=&\frac{1}{4|f|^{2}}(2|f|^{2}\nabla\overline{f}A\nabla f- f^{2}\nabla\overline{f} A\nabla\overline{f}-\overline{f}^{2}\nabla fA \nabla f)\\[2mm]
=&\frac{1}{|f|^{2}}({\rm Re}f\nabla{\rm Im}f- {\rm Im}f\nabla{\rm Re}f)A({\rm Re}f\nabla{\rm Im}f- {\rm Im}f\nabla{\rm Re}f)\geq 0.
\end{array}
\]
Hence the first line in the rhs of (\ref{oro5}) is positive.
Concerning the third line, we recall that (\ref{oro11}) implies that
$w\geq 0$ if $\alpha= \alpha_{0}a$. Since $\chi\geq 0$ the third line is also positive.  Therefore:
\[
\begin{array}{rl}
{\rm Re}\langle f-f_{\Lambda}, h_{\psi}^{*}f\rangle\geq &(\nabla \chi| A\nabla \chi)-2(F\chi| A\nabla \chi)+(\chi| c^{2}w \chi)\\[2mm]
=&\langle \chi,h_{\psi}^{*}\chi\rangle ={\rm Re}\langle \chi, h_{\psi}^{*}\chi\rangle,
\end{array}
\]
using (\ref{oro6}) and the fact that $\chi$ is real.  Using (\ref{oro7}) we obtain condition (\ref{criterion}). This completes the proof of Thm. \ref{heatkernelbound}. \qed

\subsection{Lower bounds for differential operators}\label{sec1.3}
We now deduce lower bounds for powers of $h$ from the heat kernel
bounds in Subsect. \ref{sec1.2}.
\begin{theoreme}\label{urgh}
Assume hypotheses (\ref{e1.1}), (\ref{e1.1bis}) and 
\[
v(x)\geq a \x^{-2}, \ a> 0.
\]
Then
\[
h^{-\beta}\leq C \x^{2\beta + \epsilon}, \forall \ 0\leq \beta\leq
d/2, \ \epsilon>0.
\]
\end{theoreme}
We start  by an easy consequence of Sobolev inequality.
\begin{lemma}\label{ild}
 On $L^{2}(\rr^{d})$ the following inequality holds:
 \[
(-\Delta)^{- \gamma}\leq C\x^{2\delta}, \ \forall \ 0\leq
\gamma<d/2, \ \delta>\gamma.
\]
\end{lemma}
\proof
We have
\[
(f|(-\Delta)^{-\gamma}f)= C\int\int
\frac{ \overline{f}(x) f(y)}{|x-y|^{d-2\gamma}}\d x \d y, \ \forall \ 
0<\gamma< n/2.
 \]
 By the Sobolev inequality (\cite[Equ. IX.19]{RS}):
 \[
\int\int
\frac{ \overline{f}(x) f(y)}{|x-y|^{d-2\gamma}}\d x \d y\leq C\|f\|^{2}_{r},
\]
for $r=2d/(d+ 2\gamma)$. We write then $f=
\x^{-\alpha}\x^{\alpha}f$ and use H\"{o}lder inequality to get:
\[
\|f\|_{r}\leq \|\x^{-\alpha}\|_{p}\|\x^{\alpha}f\|_{q}, \ p^{-1}+ q^{-1}= r^{-1}.
\]
We choose $q=2$, $p= d/\gamma$.  The function $\x^{-\alpha}$ belongs
to $L^{d/\gamma}$ if $\alpha>\gamma$. This implies the lemma. \qed

{\bf Proof of Thm. \ref{urgh}.}

We first recall the formula:
\begin{equation}
\label{laplace}
\lambda^{-1-\nu}= \frac{1}{\Gamma(\nu+1)}\int_{0}^{+\infty}\e^{-t
\lambda} t^{\nu}\d t, \ \nu>-1.
\end{equation}
In the estimates below, various quantities like $(f|h^{-\delta}f)$ appear. To avoid domain questions, it suffices to
replace $h$ by $h+m$, $m>0$, obtaining estimates uniform in $m$ 
and letting $m\to 0$ at the end of the proof.
We will hence prove the bounds
\beq\label{oro10}
(f| (h+m)^{-\beta}f)\leq C (f|\x^{2\beta+\epsilon}f), \ \forall \ f\in \coinf(\rr^{d}),
\eeq
uniformly in $m>0$. Moreover we note that it suffices to prove (\ref{oro10}) for $f\geq 0$. 
In fact it follows from (\ref{laplace}) that $(h+m)^{-\beta}$ has a positive kernel. 
Therefore
\[
(f|(h+m)^{-\beta}f)\leq (|f||(h+m)^{\beta}|f|)\leq C (|f||
\x^{2\beta+\epsilon}|f|)= C(f|\x^{2\beta+\epsilon}f),
\]
and (\ref{oro10}) extends to all $f\in \coinf(\rr^{d})$.

We will use the  bound (\ref{oro12}) in Thm. \ref{main}, noting that
if (\ref{oro12}) holds for some $\alpha_{0}>0$ it holds also for all
$0<\alpha\leq \alpha_{0}$. We use  the inequality
\[
\left(\frac{\x^{2}}{\x^{2}+t}\right)\left(\frac{\y^{2}}{\y^{2}+t}\right)\leq \frac{\y^{2}}{t},
\]
and get  for $f\in \coinf(\rr^{d})$, $f\geq 0$:
\[
\begin{array}{rl}
h^{-\beta}f(x)=&c\int_{0}^{+\infty}t^{\beta-1}\e^{- th}f(x)\d t\\[2mm]
\leq &C\int_{0}^{+\infty}t^{\beta-\alpha-1}(\e^{ ct\Delta}\x^{2\alpha})f(x)\d t\\[2mm]
=& C'(-\Delta)^{\beta-\alpha}\x^{2\alpha})f(x),
\end{array}
\]
as long as $\beta>\alpha$,
using again (\ref{laplace}). Integrating this point-wise inequality, we
get that
\[
(f|h^{-2\beta}f)\leq
C(f|\x^{2\alpha}(-\Delta)^{-2(\beta-\alpha)}\x^{2\alpha}f).
\]
We can apply Lemma \ref{ild} as long as $2(\beta-\alpha)< d/2$, and
obtain
\[
(f|h^{-2\beta}f)\leq
C(f|\x^{4\beta+\epsilon}f), \ \forall \  \epsilon>0,
\]
if $\alpha<\beta <\alpha+ d/4$. Since $\alpha$ can be taken
arbitrarily close to $0$, this completes the proof of the theorem.
\qed

\end{document}